\documentclass[fleqn,showpacs,twocolumn,amsmath,amssymb]{revtex4-1}

\usepackage{forloop,ifthen} 
\usepackage{multirow} 
\usepackage{ esint }

\usepackage{amsmath} 

    \usepackage{graphicx}
    \usepackage{transparent}
    \usepackage{fancybox}  
    \usepackage{color}
    \usepackage{dcolumn}
    \usepackage{bm}
    \usepackage{xspace}
    \usepackage[colorlinks=true,     linkcolor={red!70!black},
    citecolor={green!50!black}, urlcolor={blue!50!black}, pdfborder={0 0 0}]{hyperref}

\usepackage{calc}

\usepackage[usenames,dvipsnames]{xcolor}
\usepackage{ amssymb }

\newcommand{\suppress}[1]{}

\newcommand{\emphCaption}[1]{{\bf{#1}}}
\newcommand{\bea}[1]{\begin{eqnarray}\label{#1}}
\newcommand{\eea}{\end{eqnarray}}

\begin{document}

\title{Wigner's representation of quantum mechanics in integral form and its applications}

\author{Dimitris Kakofengitis}
\author{Maxime Oliva}
\author{Ole Steuernagel}

\affiliation{School of Physics, Astronomy and Mathematics, University of Hertfordshire, Hatfield,
  AL10 9AB, UK}
\date{\today}

\begin{abstract}
  We consider quantum phase-space dynamics using Wigner's representation of quantum
  mechanics. We stress the usefulness of the integral form for the description of Wigner's
  phase-space current~$\bm J$ as an alternative to the popular Moyal bracket. The integral
  form brings out the symmetries between momentum and position representations of quantum
  mechanics, is numerically stable, and allows us to perform some calculations using
  elementary integrals instead of Groenewold star products. Our central result is an
  explicit, elementary proof which shows that only systems up to quadratic in their
  potential fulfill Liouville's theorem of volume preservation in quantum mechanics.
  Contrary to a recent suggestion, our proof shows that the non-Liouvillian character of
  quantum phase-space dynamics cannot be transformed away.
\end{abstract}

\maketitle

\section{Motivation and Introduction \label{sec_intro}}

Wigner's representation of quantum mechanics in phase-space~\cite{Wigner_PR32} is
equivalent to Heisenberg's, Schr\"odinger's and Feynman's~\cite{Zachos_book_05}. The
description of the time evolution of Wigner's phase-space distribution function~$W$ uses
Moyal brackets~\cite{Moyal_MPCPS49}, the quantum analog of classical Poisson brackets. The
similarity of the Moyal form with classical physics explains its popularity. 

Moyal's bracket is defined as an infinite series of derivatives, which can make it
cumbersome to use and also numerically unstable.  It has limited applications because it
assumes that the potential can be Taylor expanded.  The integral form of quantum
phase-space dynamics~\cite{Wigner_PR32,Baker_PR58} is an alternative to Moyal's form, it
also applies to piecewise or singular potentials and displays symmetries between momentum
and position representation not obvious when using Moyal's formulation only.

We recently showed that in anharmonic quantum systems the violation of Liouville's volume
preservation can be so large that quantum phase-space volumes locally change at singular
rates~\cite{Oliva_Traj1611}. These singularities are of central importance; they
are responsible for the generation of quantum coherences.

Here, we investigate a recent suggestion by Daligault~\cite{Daligault_PRA03}, who provided
a recipe that might enable us to ``transform away'' the violation of Liouville's theorem in
anharmonic quantum-mechanical systems.

We illustrate the power of the integral form of Wigner's representation, which allows us 
to give an elementary proof that Daligault's suggestion amounts to a specific modification
that makes the dynamics classical and is incompatible with his stated aim of finding a
Liouvillian system that reproduces quantum dynamics. 

Our proof shows that the singularities, reported in
Ref.~\cite{Oliva_Traj1611}, cannot be removed to make quantum phase-space
dynamics divergence-free.

\section{Wigner's distribution and its evolution\label{sec:_WdistEvolution}}

In Wigner's representation of quantum mechanics~\cite{Wigner_PR32,Hillery_PR84} Wigner's
phase-space distribution is the \emph{``closest quantum analogue of the classical
  phase-space distribution''}~\cite{Zurek_01}. It is defined as
\begin{eqnarray}
  \label{eq:Wigner_Function1}
  W(x,p,t) & =\frac{1}{\pi\hbar}\int dy \; \varrho(x-y,x+y,t)e^{\frac{2i}{\hbar}py}\; ,  \quad
\\
  \label{eq:Wigner_Function2}
  & =\frac{1}{\pi\hbar}\int  ds\; \tilde\varrho(p-s,p+s,t)e^{-\frac{2i}{\hbar}xs}\, ; \quad
 \end{eqnarray}
 here, $\hbar=h/(2\pi)$ is Planck's constant, integrals run from~$-\infty$ to $+\infty$:
 $\int = \int_{-\infty}^{\infty}$, and $\varrho$ and~$\tilde \varrho$ are the density
 operator in position and momentum representation, respectively.

Wigner's distribution~$W$ is set apart from other quantum phase-space
distributions~\cite{Hillery_PR84} by the fact that only~$W$ simultaneously yields the
correct projections in position and momentum ($\varrho(x,x,t)= \int
dp \; W$ and $\tilde \varrho(p,p,t)=\int
dx \; W$) as well as state overlaps $|\langle \psi_1 | \psi_2 \rangle |^2 = 2 \pi \hbar
\int
\int
dx\; dp \; W_1 \; W_2$, while maintaining its form~(\ref{eq:Wigner_Function1}) when evolved in
time. Additionally, the Wigner distribution's averages and uncertainties evolve
\emph{momentarily} classically~\cite{Royer_FOP92,Ballentine_PRA94}.

In this work we consider one-dimensional conservative systems in a pure state with
quantum-\emph{mechanical} Hamiltonians
\begin{equation}
  \label{eq:Classical_Hamiltonian}
  {\cal \hat H}(\hat x,\hat p) = \frac{\hat p^2}{2M} + \hat V(\hat x) \; .
\end{equation}
The Wigner function's time evolution arises, in analogy to
Eq.~(\ref{eq:Wigner_Function1}), from a Wigner-transform [which can be implemented as a
fast Fourier transform (FFT)] of the von Neumann equation $ i\hbar\frac{\partial \hat\varrho}{\partial t} = [{\cal
  \hat H},\hat\varrho]$ as~\cite{Wigner_PR32,Hillery_PR84}
\begin{flalign}
  \label{eq:Quantum_Liouville_Integral_Form}
  \partial_t W =& -\frac{p}{M}
  \frac{1}{\pi\hbar}\int
 dy \, \partial_x\varrho(x-y,x+y,t)e^{\frac{2i}{\hbar}py} \nonumber \\
  &+ \frac{i}{\pi\hbar^2}\int
 dy \; \left[V(x+y)-V(x-y)\right] \nonumber \\
  & \quad \times\varrho(x-y,x+y,t) e^{\frac{2i}{\hbar}py}\, ,
\end{flalign}
also known as the quantum Liouville equation. 

Throughout, we write partial derivatives as~
$\frac{\partial^n}{\partial x^n}=\partial_x^n$.

If the potential~$V$ can be globally Taylor expanded, the
integrals~(\ref{eq:Quantum_Liouville_Integral_Form}) yields the Moyal bracket
$\left\{\left\{\cdot,\cdot\right\}\right\}$~\cite{Moyal_MPCPS49}
\begin{flalign}
  \label{eq:Wigner_Function_Time_Evolution1}
  \frac{\partial W}{\partial t} &= \left\{\left\{ {\cal H},W \right\}\right\} = \frac{1}{i \hbar} \left( {\cal H} \star W - W \star {\cal H} \right) \\
  \label{eq:Wigner_Function_Time_Evolution2}
  &= \frac{2}{\hbar} {\cal H} \sin\left( \frac{\hbar}{2} \left( \overleftarrow{\partial_x}
      \overrightarrow{\partial_p} - \overleftarrow{\partial_p} \overrightarrow{\partial_x} \right)
  \right) W \; .
\end{flalign}
Here we use Groenewold star products ($\star$)~\cite{Groenewold_Phys46}, defined as $f
\star g = f e^{\frac{i \hbar}{2}\left( \overleftarrow{\partial_x}
    \overrightarrow{\partial_p} - \overleftarrow{\partial_p} \overrightarrow{\partial_x}
  \right)} g$; the arrows indicate whether derivatives are executed on~$f$ or~$g$.

Equations~(\ref{eq:Quantum_Liouville_Integral_Form})
or~(\ref{eq:Wigner_Function_Time_Evolution2}) can be written as the continuity
equation~\cite{Wigner_PR32}
\begin{equation}
  \label{eq:Continuity_Equation}
  \partial_t W + \bm{\nabla} \cdot \bm{J} = \partial_t W + \partial_x J_x + \partial_p J_p = 0\; .
\end{equation}

Comparing Eqs.~(\ref{eq:Continuity_Equation}) and~(\ref{eq:Quantum_Liouville_Integral_Form}), we
identify the Wigner current~$\bm J =\binom{ J_x }{ J_p }$, with position component
\begin{flalign}
 & \;\; J_x  = \frac{p}{M\pi\hbar}\int
  dy\; \varrho(x-y,x+y,t)e^{\frac{2i}{\hbar}py} = \frac{p}{M} W \, ,&
  \label{eq:Wigner_Current_Jx}
\end{flalign}
and momentum component 
\begin{flalign}
  J_p  = & -\frac{1}{\pi\hbar}\int
  dy\; \left[\frac{V(x+y)-V(x-y)}{2y}\right]
  \nonumber\\
  \times &
  \varrho(x-y,x+y,t) e^{\frac{2i}{\hbar}py}\, .
 \label{eq:Wigner_Current_Jp_Integral_Form}
 \end{flalign}
 If the potential can be Taylor expanded, the explicit form of the components of Wigner
 current~$\bm J$ in Eq.~(\ref{eq:Wigner_Function_Time_Evolution2})
 is~\cite{Wigner_PR32,Skodje_PRA89,Donoso_PRL01,Ole_PRL13}
\begin{flalign} {\bm J} = \bm j +
 \begin{pmatrix} 0
    \\-\sum\limits_{l=1}^{\infty}{\frac{(i\hbar/2)^{2l}}{(2l+1)!}
      \partial_p^{2l}W \partial_x^{2l+1}V(x) }
 \end{pmatrix}
\; . \quad
\label{eq:CurrentComponentsMT}
\end{flalign}
Here, with $\bm v = \binom{ {p}/{M} }{- \partial_x V }$, $\bm j = W \bm v$ is the
classical and $\bm J - \bm j$ are quantum terms.

\section{Features and applications of the integral form\label{sec:_FeaturesApplicationsIntegralForm}}

The integral form~(\ref{eq:Quantum_Liouville_Integral_Form}) is more general than the Moyal
expression since it does not rely on $V$ being analytic.

A numerical implementation of the integral form can use fast Fourier transforms. In the
case of potentials featuring high order Taylor terms, high order numerical derivatives can
render Eq.~(\ref{eq:CurrentComponentsMT}) poorly convergent~\cite{Ole_PRL13}.

In Ref.~\cite{Ole_PRL13} we showed (note typographical errors in~Eqs.~(4) and~(5)
of~\cite{Ole_PRL13}) that the $p$~projection of~$J_x$ yields the quantum probability
 current~$\jmath$ in position space,
\begin{flalign}
  \int   \, dp & J_x  =  \frac{\hbar}{2iM} \int
 \, dy \, \varrho(x-y,x+y,t)
  \partial_y\delta(y) \nonumber &\\
  = & \sum_{k} P_k \frac{\hbar}{2iM} \left(\Psi_k^*\partial_x \Psi_k - \Psi_k \partial_x
    \Psi_k^*\right) = \jmath(x,t) \; ,&
 \label{eq:Probability_Current_Position_Space}
\end{flalign}
where we used Dirac's $\delta$ and wrote the density matrix as $ \varrho (x,x',t) =
\sum_{k} P_k \Psi_k (x,t) \Psi_k^* (x',t)$, a statistical mixture of pure
states. Additionally,
\begin{flalign}
  \int
  & \, dx J_x 
  = \frac{p}{M} \int
  \, ds \, \tilde\varrho(p-s,p+s,t) \delta(s) = \frac{p}{M}\tilde\varrho(p,p,t)\; .&
  \label{eq:Wigner_Current_Jx_x_projection}
\end{flalign}
Analogously to Eq.~(\ref{eq:Probability_Current_Position_Space}), the quantum probability
current~$\tilde\jmath$ in momentum space~\cite{Ole_PRL13}, is
\begin{flalign}
  \int dx \; J_p &= -\frac{1}{\pi\hbar} \iint
  \, dy\, dx \,
  \left[\frac{V(x+y)-V(x-y)}{2y}\right] \nonumber&\\
  &\quad\times \varrho(x-y,x+y,t)e^{\frac{2i}{\hbar}py} \nonumber&\\ &=\frac{1}{i\sqrt{2\pi\hbar^3}}
  \int_{-\infty}^p dp'\int
  dp''\; \tilde V^*(p''-p')\tilde\varrho(p'',p') \nonumber&\\
  &\quad-\tilde V(p''-p')\tilde\varrho(p',p'')= \tilde\jmath (p,t)\; ,&
  \label{eq:Probability_Current_Momentum_Space}
\end{flalign}
(where~$\tilde V (p)=\frac{1}{\sqrt{2\pi\hbar}}\int dx\; V(x)e^{-\frac{i}{\hbar}px}$), while
\begin{flalign}
  \int
  \, dp \, J_p = & -\int
  \, dy \, \left[\frac{V(x+y)-V(x-y)}{2y}\right] \nonumber \\ & \times\varrho(x-y,x+y,t)\delta(y)
  \nonumber \\
  &= -\varrho(x,x,t) \frac{dV}{dx} \; .
  \label{eq:Wigner_Current_Jp_p_projection1}
\end{flalign}
We would like to emphasize that the quantum terms of Eq.~(\ref{eq:CurrentComponentsMT}) do not contribute in
Eq.~(\ref{eq:Wigner_Current_Jp_p_projection1}). 

Averaging over Eqs.~(\ref{eq:Wigner_Current_Jx_x_projection}) and~(\ref{eq:Wigner_Current_Jp_p_projection1}),
reproduces Ehrenfest's theorem~\cite{Zachos_book_05}
\begin{flalign}
  \label{eq:Ehrenfest_Theorem_Average_velocity}
  \iint dx dp \; J_x = \frac{\left\langle \hat p\right\rangle}{M} = \frac{d\langle \hat
    x\rangle}{dt} &
\end{flalign}
and
\begin{flalign}
  \label{eq:Ehrenfest_Theorem_Average_Force}
  \iint dx dp \; J_p = -\left\langle \frac{d \hat V}{dx} \right\rangle = \frac{d\langle
    \hat p\rangle}{dt}\, .&
\end{flalign}
Where applicable, the Moyal bracket formalism~\cite{Zachos_book_05} yields the same
results.  Note the various subtleties associated with the interpretation of Ehrenfest's
theorem~\cite{Ballentine_PRA94,Royer_FOP92}.

\section{When is quantum mechanical time evolution Liouvillian? \label{sec_Liouvillian}}

To investigate whether quantum phase-space dynamics is Liouvillian we determine the
divergence of its quantum phase-space velocity field~$\bm
w$~\cite{Donoso_PRL01,Trahan_JCP03,Daligault_PRA03}. $\bm w$ is the quantum analog of the
classical velocity field~$\bm v$~(Eq.~(\ref{eq:CurrentComponentsMT})):
\begin{flalign}
 \label{eq:Wigner_w_velocityMT}
\bm w = & \frac{\bm J}{W} = \bm v + \frac{1}{W}
 \begin{pmatrix} 0
    \\-\sum\limits_{l=1}^{\infty}{\frac{(i\hbar/2)^{2l}}{(2l+1)!}
      \partial_p^{2l}W \partial_x^{2l+1}V } 
 \end{pmatrix}. &
\end{flalign}
To rephrase the continuity equation~(\ref{eq:Continuity_Equation}) in terms of~$\bm w$, we
switch to the Lagrangian decomposition~\cite{Donoso_PRL01,Trahan_JCP03,Daligault_PRA03}
\begin{equation} 
  \frac{d W}{d t} = \partial_t W + \bm w \cdot \bm \nabla W =- W \bm \nabla \cdot \bm w \; . 
\label{eq:W_TotalDerivMT}
\end{equation}

Note that $\bm w$ is singular at zeros of~$W$ since, generally, zeros of~$W$ do not
coincide with zeros of its derivatives. This implies, among other things, that the
concept of trajectories in quantum phase-space cannot be applied to the dynamics of
anharmonic systems~\cite{Oliva_Traj1611}.

Problems associated with the singularities have been observed multiple
times~\cite{Sala_JCP93,Trahan_JCP03}, they badly affect numerical quantum phase-space
studies~\cite{Trahan_JCP03}.

It would therefore be intriguing to be able to transform such problems away, as suggested
by Daligault~\cite{Daligault_PRA03}. He speculated that it might be possible to add an
auxiliary field~$\delta \bm J$ to $\bm J$ in Eq.~(\ref{eq:Continuity_Equation}) which
would not modify the dynamics since it is assumed to be divergence-free. Yet, this
auxiliary field might yield a modification to the velocity field such that their sum
fulfills Liouville's theorem:~$\bm \nabla \cdot (\bm w + \delta \bm w)=0.$ If possible, we
could deploy the machinery of classical phase-space transport equations to solve quantum
problems.

We now prove that we cannot get rid of the non-Liouvillian character of quantum
phase-space dynamics in anharmonic systems in the way Daligault suggested.

  To do this, we need to establish when~$\bm J$ obeys Liouville's theorem, i.e., when the
  divergence of $\bm J$'s velocity field vanishes everywhere in phase-space. With $\bm
  w = \bm J /W$ we have
\begin{flalign}
  \bm{\nabla}\cdot\bm{w} & = \partial_x\left(\frac{J_x}{W}\right) +
  \partial_p\left(\frac{J_p}{W}\right)
 = \partial_x\left(\frac{p}{M}\right)
  + \partial_p\left(\frac{J_p}{W}\right) & \nonumber \\
& = \partial_p\left(\frac{J_p}{W}\right) = 0. &
  \label{eq:WignerCurrent_VelocityDivergence2}
\end{flalign}
Integrating gives us $\int_{-\infty}^p dp' \;\partial_{p'} \frac{J_{p}(x,p',t)}{W} =
\frac{J_p}{W} = C(x) $ which implies $ \int
dp \; J_p = C(x)\int dp \; W = C(x) \; \varrho(x,x,t).$ With $- \frac{d V}{d x}
\;\varrho(x,x,t) = C(x) \; \varrho(x,x,t)$,
from~(\ref{eq:Wigner_Current_Jp_p_projection1}), it follows that~$\bm \nabla \cdot \bm w
=0$ implies
\begin{equation}
  \label{eq:WignerCurrent_VelocityDivergence004}
  J_p + \delta J_p = -W \; \frac{\partial}{\partial x} [V(x) + \delta V(x)]\; .
\end{equation}
We have shown that the application of Daligault's recipe filters through in a very
specific form: the dynamics becomes classical and the shift only affects the potential
(since the goal is to not affect the time evolution of~$W$). Strictly speaking, in
Eq.~(\ref{eq:WignerCurrent_VelocityDivergence004}), we should write~$\delta V(x,W(x,p,t))$
to remind ourselves of Daligault's assumption that~$\delta J_p$ depends on~$W$. But to
yield Liouvillian dynamics $\delta V$ must not depend on~$p$; hence~$\delta
V(x,W(x,p,t))=\delta V(x)$.

For systems in which the potential can be globally Taylor expanded,
Eq.~(\ref{eq:WignerCurrent_VelocityDivergence004}) shows us that quantum terms must not be
present in~Eq.~(\ref{eq:CurrentComponentsMT}). To fulfill Liouvillian behavior for all
times the potential~$V$ might be of ``harmonic'' form:~$V = V_\text{harmonic} =
\frac{K}{2} x^2 + a x +b$ with arbitrary real $K$, $a$, and $b$ and, therefore,~$\delta
V=0$.

Alternatively, the auxiliary field has the trivial form $\delta \bm J = \bm j - \bm J$
which subtracts all the quantum parts in~Eq.~(\ref{eq:CurrentComponentsMT}), so that the
potential assumes the form~$V+\delta V = V_\text{harmonic}$. This is neither what Daligault
intended nor is it helpful, in fact, it is not even permissible since such a field would
not fulfill the condition $\bm \nabla \cdot \delta \bm J = 0$.

One might wonder whether there is some other option, perhaps the anharmonic quantum
terms~$\bm J - \bm j$ in Eq.~(\ref{eq:CurrentComponentsMT}) are present but the initial
state~$W_0$ has some special form that cancels all anharmonic terms yet does not force the
trivial form $V+\delta V = V_\text{harmonic}$ on us.

This cannot be though: if $J_p + \delta J_p$ fulfills
Eq.~(\ref{eq:WignerCurrent_VelocityDivergence004}), the dynamics is classical and
anharmonic, shearing the phase-space distribution.  Since a Wigner distribution can be
expanded in the coherent-state basis, we assume, without loss of generality, that the
initial state~$W_0$ is a coherent state. Classically shearing a phase-space distribution
bends it out of shape while keeping it positive, this violates the constraint that a
positive Wigner distribution has Gaussian form~\cite{Hudson_RMP74}: anharmonic
positivity-preserving classical dynamics is incompatible with quantum phase-space
dynamics.

Generalizing Daligault's recipe slightly: might modifications to the~$J_x$ component help?
We doubt it, if the system is Hamiltonian; even if not quantum \emph{mechanical} but, say,
of the classical Kerr-oscillator type, one would, according to
Eq.~(\ref{eq:W_TotalDerivMT}), still end up with Liouvillian dynamics:~$\frac{d}{dt} W
=0$. $W$ cannot change sign, Daligault's recipe could never give us quantum dynamics,
i.e., negativity formation in phase-space~\cite{Oliva_Traj1611}.

\begin{figure*}[t]
\centering
\includegraphics[width=148mm,angle=0]{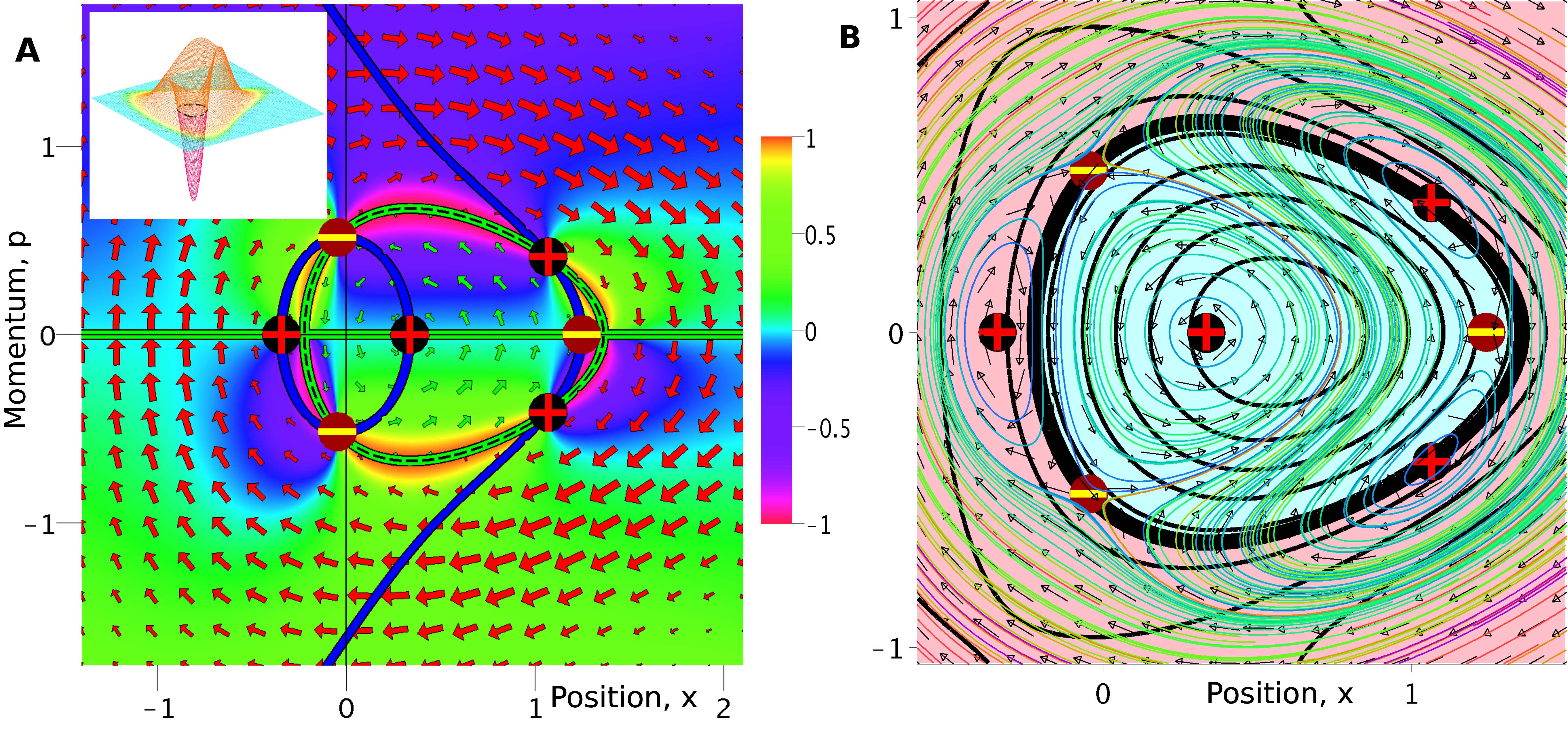}
  \caption{\textsf{\textbf{A}}, \emphCaption{Singularities of~$\bm \nabla \cdot \bm w$
      coincide with zeros of~$W$.} ${\bm J}$ depicted by arrows (red for clockwise and
    green for inverted flow~\cite{Ole_PRL13}), together with the zeros of the $J_x$ and
    $J_p$ components (green and blue lines, respectively), is superimposed on top of a
    colorplot of $\frac{2}{\pi} \arctan (\bm \nabla \cdot \bm w)$. The inset shows the
    corresponding Wigner distribution for the first excited state of an anharmonic Morse
    oscillator~\cite{Dahl_JCP88} with
    potential~$V(x)=3[1-\exp({{-x}/{\sqrt{6}}})]^2$. The red crosses and yellow bars
    mark the locations of the flow's stagnation points, with Poincar\'e-Hopf
    indices~\cite{Ole_PRL13}~$\omega=+1$ and~$-1$. Parameters:~$\hbar=1$ and $M=1$. The black
    dashed line marks the zero contour of the Wigner distributions (compare inset); here 
the divergence~$\bm \nabla \cdot \bm w$ is singular~\cite{Oliva_Traj1611}.  
\newline \textsf{\textbf{B}},
    \emphCaption{Integrated Fieldlines of $\bm J$ cross Wigner Distribution
      Contours}. Thin colored lines display fieldlines of $\bm J$, displayed together
    with normalized current ${\bm J}/||\bm J||$ (black arrows), and its stagnation points, for the
    same state as depicted in~\textsf{\textbf{A}}. $W$'s zero contour, around the
    negative (light cyan-colored) patch at the center, is highlighted by a thick black
    line. Many fieldlines, for this first excited state, cut across the Wigner
    distribution's contours and enter and leave the negative area.
    \label{fig:divergence}}
\end{figure*}

In their monograph on the Wigner representation, Zachos \emph{et
  al.}~\cite{Zachos_book_05} argue that anharmonic quantum systems cannot fulfill
Liouville's theorem since the difference between the Moyal and Poisson brackets is
nonzero for anharmonic quantum systems. In light of Daligault's speculation that a
mapping to \emph{another} system might exist, that reproduces the same dynamics and
fulfills Liouville's theorem, we feel the above proof with the explicit use of~$\bm J$ is
needed to settle the matter.

Figure~\ref{fig:divergence} shows that the divergence~$\bm \nabla \cdot \bm w$ becomes
singular when $W=0$. This follows from Eq.~(\ref{eq:Wigner_w_velocityMT}). It indicates
qualitatively that, since quantum states almost always have
zero-lines~\cite{Hudson_RMP74}, there will almost always be regions of singularities
of~$\bm \nabla \cdot \bm w$. In this sense, the attempt to transform away non-Liouvillian
behavior of quantum dynamics appears futile. Instead, such divergences explain certain
numerical problems~\cite{Oliva_Traj1611} and emphasize how very different quantum and
classical phase-space dynamics are: whereas classical dynamics constitute one extreme
($\bm \nabla \cdot \bm w =0$ always), quantum dynamics (for anharmonic systems) constitute
another ($|\bm \nabla \cdot \bm w | = \infty$, almost always, somewhere in phase-space).

Since Eq.~(\ref{eq:Continuity_Equation}) features singular divergence of the velocity
field one should perhaps stress quantum dynamics from the continuity equation point of
view rather than referring to quantum Liouville equations.

\section{Conclusion \label{sec_Conclusion}}

We showed that the integral form of Wigner's representation of quantum mechanics should be
consulted as an alternative to Moyal's formulation. It is more general than Moyal's
form. If high order derivatives are present in Moyal's form, the integral form tends to
converge better in numerical calculations, which can be implemented as FFTs. It can make
mathematical manipulations more transparent than Moyal's form, and it displays symmetries
between position and momentum configuration space more clearly.

\end{document}